\documentclass[aps,pre,twocolumn,showpacs,superscriptaddress,reprint,longbibliography]{revtex4-1}
\usepackage[english]{babel}
\usepackage[utf8]{inputenc}
\usepackage{amsmath}
\usepackage{amssymb}
\usepackage{graphicx}
\usepackage[normalem]{ulem}
\usepackage{hyperref}
\usepackage{epstopdf}
\usepackage{color}

\def\ddt{\partial_t}
\def\ddx{\partial_x}

\newcommand{\vk}{\mathbf{k}}

\def\bea{\begin{eqnarray}}
\def\eea{\end{eqnarray}}

\begin{document}
\title{Radiation-Pressure-Driven Ion Weibel Instability and Collisionless Shocks}

\author{A. Grassi}
\email[Corresponding author: ]{a.grassi8@gmail.com }
\affiliation{LULI, Sorbonne Université, CNRS, Ecole Polytechnique, CEA, Université Paris-Saclay, Paris, France}
\affiliation{Dipartimento di Fisica Enrico Fermi, Università di Pisa, Largo Bruno Pontecorvo 3, I-56127 Pisa, Italy}
\affiliation{Istituto Nazionale di Ottica, Consiglio Nazionale delle Ricerche (CNR/INO), u.o.s. Adriano Gozzini, I-56127 Pisa, Italy}

\author{M. Grech}
\affiliation{LULI, CNRS, Ecole Polytechnique, CEA, Université Paris-Saclay, Sorbonne Université, Palaiseau, France}

\author{F. Amiranoff}
\affiliation{LULI, Sorbonne Université, CNRS, Ecole Polytechnique, CEA, Université Paris-Saclay, Paris, France}

\author{A. Macchi}
\affiliation{Dipartimento di Fisica Enrico Fermi, Università di Pisa, Largo Bruno Pontecorvo 3, I-56127 Pisa, Italy}
\affiliation{Istituto Nazionale di Ottica, Consiglio Nazionale delle Ricerche (CNR/INO), u.o.s. Adriano Gozzini, I-56127 Pisa, Italy}

\author{C. Riconda}
\affiliation{LULI, Sorbonne Université, CNRS, Ecole Polytechnique, CEA, Université Paris-Saclay, Paris, France}

\date{\today}

\begin{abstract}
The Weibel instability from counterstreaming plasma flows is a basic process highly relevant for collisionless shock formation in astrophysics. In this paper we investigate, via two- and three-dimensional simulations, suitable configurations for laboratory investigations of the ion Weibel instability (IWI) driven by a fast quasi-neutral plasma flow launched into the target via the radiation pressure of an ultra-high-intensity (UHI) laser pulse ("Hole-Boring" process). The use of S-polarized light at oblique incidence is found to be an optimal configuration for driving IWI, as it prevents the development of surface rippling observed at normal incidence, that would lead to strong electron heating and would favor competing instabilities. Conditions for the evolution of IWI into a collisionless shock are also investigated.
\end{abstract}

\maketitle

%%% INTRODUCTION
\section{Introduction}
Today's high-intensity laser facilities open new possibilities for the study, in the laboratory, of scenarios relevant to various astrophysical processes, 
among which collisionless shocks have recently attracted remarkable interest~\cite{Zakharov2003,Ryutov2012,FiuzaPRL,Stockem2014,Higginson2015,Huntington2015,Ruyer2015,Lobet2015}.
Collisionless shocks are ubiquitous in a wide range of astrophysical environments (active galaxy nuclei, pulsar wind nebulae, supernovae remnants, etc.).
They develop in the presence of fast counter-streaming plasma flows and are held responsible for non-thermal particles in cosmic rays and high-energy radiation~\cite{Kirk1999}.

In the absence of particle collisions, the dissipation of the flow kinetic energy into thermal energy necessary to shock formation is mediated by micro-instabilities. 
Among these, the Weibel instability~\cite{Weibel1959,GremilletPOP2010,Grassi2017} has been identified as responsible for shock formation in various environments~\cite{Medvedev1999}. It leads to the development of turbulent magnetic fluctuations that cause isotropization of the flows and, at later times, particle energetization via first order Fermi acceleration~\cite{Fermi1949}.

{\it In situ} measurements in most astrophysical systems being far beyond our reach, reproducing such non-linear processes in the laboratory would provide a unique platform for their investigation.
Therefore this line of study has attracted interest from the laser-plasma interaction community, both on the simulation~\cite{Kato2008,FiuzaPRL,Stockem2014,Ruyer2015,Lobet2015} and experimental~\cite{Huntington2015,Park2015,NIF2017} sides.
Indeed, high-intensity laser systems allow to create fast, supersonic flows which could mimic those encountered in astrophysics.

Up to now, most of the studies have focused on the use of high-energy (multi-kJ and NIF-LMJ class) laser facilities operating 
at modest intensities ($\lesssim 10^{16} {\rm W/cm^2}$)~\cite{Huntington2015,Park2015,NIF2017}. On such laser systems, 
the resulting plasma flows are created by ablation of a dense target, which limits the accessible flow density and velocity (typically 
$\lesssim 0.5\%$ of the speed of light). 
As a result, the characteristic length (few centimeters) and time (tens of nanoseconds) over which shock formation can be expected
are large. This has potentially two drawbacks. First, it requires the use of large laser systems such as NIF or LMJ.
Second, the effect of particle collisions over such lengths/times may not be completely negligible.

In contrast, ultra-high intensity (UHI) laser systems, with peak intensities beyond $10^{18}\,{\rm W/cm^2}$, could allow to alleviate these limitations 
by providing a complementary path toward the creation of collisionless, ultra-fast and high-density plasma flows.
The use of UHI laser systems to drive a collisionless shock was first proposed in Ref.~\cite{FiuzaPRL}, and further investigated in~\cite{Ruyer2015}.
In the configuration considered in Refs.~\cite{FiuzaPRL,Ruyer2015}, the physics leading to the formation of a collisionless shock is dominated
by electron instabilities driven by the interplay of laser-generated hot electrons launched into the target, and the resulting electron return current.
As a result, shock formation in this configuration follows from a process very different than that observed in astrophysical systems where ion Weibel instability dominates the shock formation process.\\

In this work, we propose a configuration where the Ion Weibel instability (IWI) can be efficiently driven by fast and dense quasi-neutral plasma flows.
In this configuration the generation of hot electrons is minimized, a situation close to the astrophysical scenarios where neutral flows of charged particles become Weibel unstable when interacting with the interstellar medium. 
The fast quasi-neutral plasma flow is created by radiation pressure in a dense target irradiated by an UHI laser beam.  
In this situation the system evolution follows a first phase in which electrons turn unstable while the large inertia of the ions keeps their trajectories weakly affected by the magnetic turbulence. On a longer timescale, IWI develops in a background of warm electrons, and the magnetic field grows up to a large-enough amplitude to efficiently deflect the ions. The progressive deceleration of the ion flow eventually produces an increase of the density that leads to the formation of a shock front~\cite{Bret2016,RuyerPRL,RuyerPOP}.
By means of Particle-In-Cell (PIC) simulations, we demonstrate that a S-polarized laser beam irradiating the target at oblique incidence is an optimal scheme to suppress strong electron heating and thus drive the IWI, and at later times an IWI-mediated collisionless shock. 
In contrast, the seemingly suitable configuration of normal incidence and circularly polarized light is shown to be affected by fast instabilities developing at
the laser-plasma interface, eventually leading to strong electron heating. As will be shown, this configuration leads to a similar situation than the one studied in Refs.~\cite{FiuzaPRL,Ruyer2015}, where the system evolution is governed by electron instabilities.

In the scheme we propose, due to the larger density and higher flow velocities (typically 10\% of the speed of light) with respect to those achievable at NIF, larger growth-rates can be obtained for the IWI. This also entails shorter times for shock formation and may help preserving the collisionless regime.\\

The paper is structured as follows. In Sec.~\ref{sec:IWI3dsim}, we present a three-dimensional (3D) PIC simulation that demonstrates how IWI can be triggered inside a dense target irradiating it with a S-polarized laser pulse at oblique ($45^{\circ}$) incidence. The characteristic filamentary structures in both current and magnetic field, as well as the instability growth rate, are discussed and confronted to an analytical model. In Sec.~\ref{sec:Shock}, we show, by means of a reduced 2D simulation (also exploiting a reduced ion to electron mass ratio for computational convenience and for this simulation only) that the radiation pressure driven IWI eventually leads to the formation of a collisionless shock. The properties of the downstream (shocked) plasma are found to be in good agreement with theoretical (Rankine-Hugoniot) predictions. In Sec~\ref{sec:SurfInsta}, we further discuss the surface instability. Using 2D simulations, we demonstrate its role in the hot electron production observed when considering an arbitrary (circularly or linearly) polarised laser pulse at normal incidence on the overdense target. The surface instability mitigation and the resulting electron heating suppression using a S-polarized light pulse at oblique ($45^{\circ}$) incidence is then explained by the creation of a strong, laser-driven surface current.
A simplified analytical model for the creation of this current sheet is proposed. Finally, Sec.~\ref{sec:conclusions} presents our conclusions.

%%% HB & IWI
\section{Radiation-pressure-driven ion Weibel instability}\label{sec:IWI3dsim}

\begin{figure}
\includegraphics[width=0.5\textwidth]{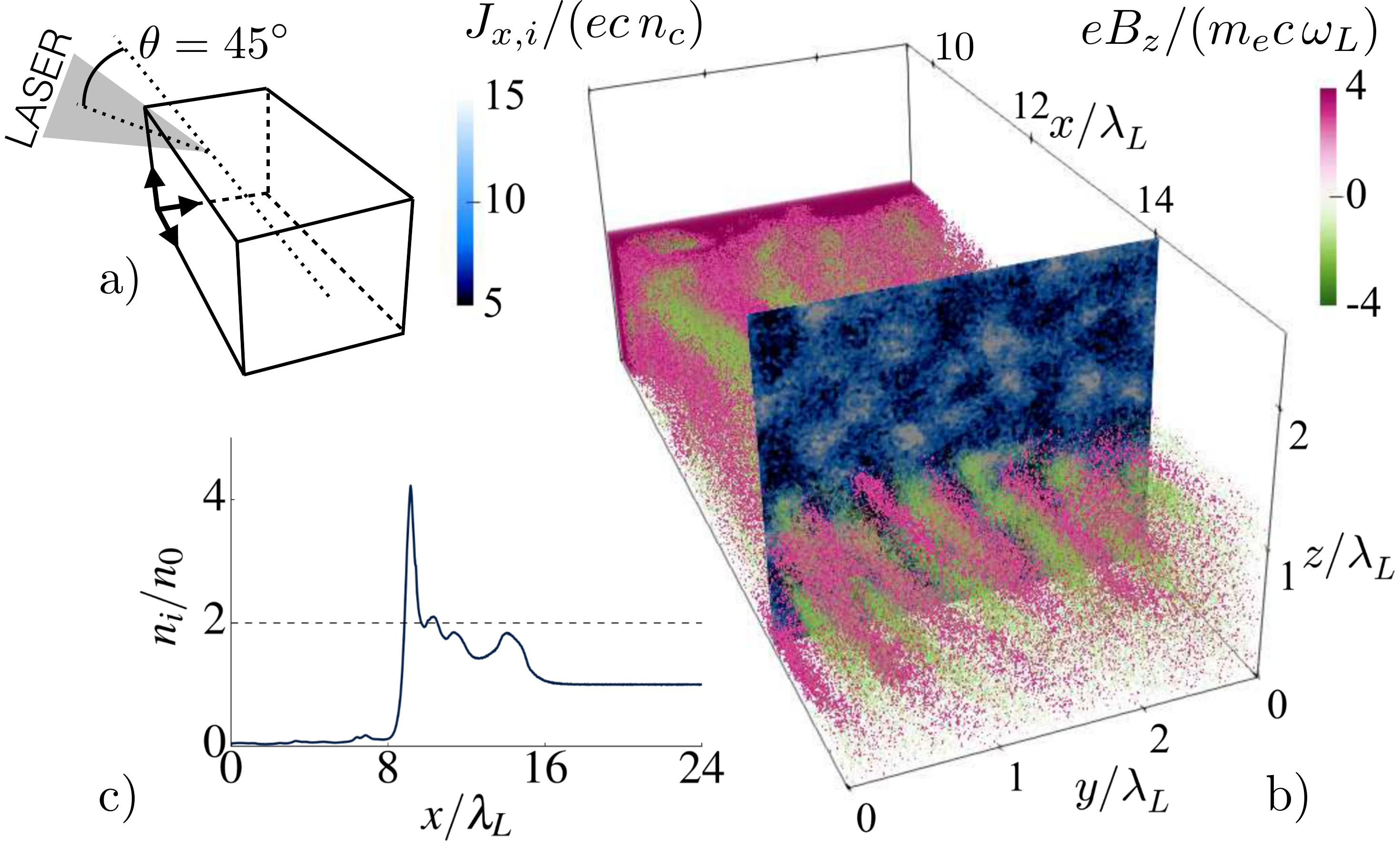}
\caption{\label{3DSpol45}a) Schematic presentation of the investigated set-up. The incident angle $\theta$ of the S-polarized laser pulse is defined in the $x$-$y$ plane. b) Ion Weibel instability from radiation pressure driven flows in a 3D PIC simulation with realistic ion mass. Magnetic field $B_z$ and slice in the $y$-$z$ plane at $x=14\,\lambda_L$ of the ion current $J_{x,i}$  at $t=65\,t_L$. c) Ion density $\langle n_i(x) \rangle_{y,z}$ averaged over the $y$-$z$ plane. The laser pulse (not shown) propagates along the $x>0$ direction, and the laser-plasma interaction surface at this time is located at $x\simeq10\,\lambda_L$.}
\end{figure}

The configuration investigated in this work is pictured in Fig.~\ref{3DSpol45}a).
In this scheme, a UHI laser pulse is incident onto an overdense target at oblique incidence.
The laser ponderomotive force pushes inward the electrons located close to the surface, 
quickly creating a double-layer structure with the ions following the electrons~\cite{Schlegel09}.
This structure acts as a piston advancing the surface at a constant velocity $v_{\mbox{\tiny HB}}$, and efficiently reflecting ions at $\simeq 2v_{\mbox{\tiny HB}}$, a process known as Hole-Boring (HB). The HB velocity is estimated by balancing the flux of ion momentum with the laser radiation pressure in the rest frame of the plasma surface and can easily reach a non-negligible fraction of the speed of light $c$. 
Assuming perfect reflection in this frame (primed quantities), the velocity is obtained by equating $P_{\mbox{\tiny rad}}'=2I'\cos^2\theta'/c=2m_in_0\gamma_{\mbox{\tiny HB}}^2v_{\mbox{\tiny HB}}^2$, where $I'=I(1-v_{\mbox{\tiny HB}}/c)/(1+v_{\mbox{\tiny HB}}/c)$ and $\theta'=\arctan^{-1}\left[\sin\theta/(\gamma_{\mbox{\tiny HB}}(\cos\theta -v_{\mbox{\tiny HB}}/c))\right]$~(see Appendix~\ref{app:Piston}); here $I$ is the laser intensity and $\theta$ the angle of incidence in the laboratory frame, $\gamma_{\mbox{\tiny HB}}=(1-v^2_{\mbox{\tiny HB}}/c^2)^{-1/2}$, $m_i$ the ion mass, and $n_0$ the unperturbed target density.
In the frame co-moving with the surface, the background plasma and the Hole-Boring reflected beam constitute two neutral counter-propagating beams with velocity $\simeq \pm v_{\mbox{\tiny HB}}$. For sufficiently high flow velocities, $v_{\mbox{\tiny HB}}\gtrsim 0.1c$, this entails a fast growth rate and ensures the Weibel instability to be the dominant mode in the unstable spectrum~\cite{GremilletPOP2010}. \\

The micro-instability at play (and the evolution of the system toward full shock formation in the next Sec.~\ref{sec:Shock}) have been investigated by means of kinetic simulations
performed with the PIC code {\sc Smilei}~\cite{SmileiPaper}. 
Figure~\ref{3DSpol45}b) shows the results of a 3D simulation carried out considering a S-polarized plane wave of intensity $I\simeq6.8\times 10^{21}\:{\rm Wcm^{-2}}/\lambda^2_{\rm \mu m}$, where $\lambda^2_{\rm \mu m}$ is the laser wavelength in units of $\rm \mu m$, corresponding to a normalized laser vector potential $a_0=70$. This electromagnetic wave irradiates, at an angle of incidence $\theta=45^{\circ}$, an electron-proton plasma ($m_i=1836\:m_e$) with density $n_0 = 49\,n_c$ ($n_c$ being the critical density) and initial (electron and ion) temperature $T=1\,{\rm keV}$. 
The extension of the simulation box is $L_x=48\,\lambda_L,\,L_y=L_z=2.5\,\lambda_L,$ where $\lambda_L$ is the laser wavelength and $x$ the direction normal to the plasma surface. The spatial resolution is $\lambda_L/64$ and 8 macroparticle-per-cell were used for each species (for a total of $\simeq 1.4\times10^9$). 
The simulation runs over $\simeq 66\:t_L$, with $t_L=\lambda_L/c$ the laser period. 
Periodic boundary conditions are used along the $y$ and $z$ directions for fields and particles.

In the simulation, the laser-plasma interaction surface is found to move at a velocity $v^{\mbox{\rm\tiny sim}}_{\mbox{\tiny HB}}\simeq0.11\,c$, in good agreement with the theoretical value of $v_{\mbox{\tiny HB}}\simeq0.10\,c$ for these parameters. 
Figure~\ref{3DSpol45}c) shows that after $65\,t_L$ the overlapping region has a density $\simeq 2n_0$. 
Figure~\ref{3DSpol45}b) demonstrates the presence of filamentary structures in both the magnetic field $B_z$ and the ion current $J_{x,i}$.
The growth rate of the magnetic energy $U_B\propto e^{2\Gamma t}$ has been measured in the simulation over a layer, in the overlapping region, with extension $\simeq0.2\,\lambda_L$, and moving at $v^{\mbox{\rm\tiny sim}}_{\mbox{\tiny HB}}$. 
We obtain $\Gamma^{\mbox{\rm\tiny sim}}_{\mbox{\rm\tiny IWI}} \simeq0.034\,t_L^{-1}$ and a dominant mode $k^{\mbox{\rm\tiny sim}}_{\mbox{{\rm\tiny IWI}}}\simeq2\,\omega_L/c$, corresponding to filamentary structures with wavelength $\lambda^{\mbox{\rm\tiny sim}}_{\mbox{{\rm\tiny IWI}}}\simeq0.5\,\lambda_L$. 
As will now be detailed, these observations are consistent with the development of the IWI. 

Relativistic fluid theory of two counter-propagating ion beams, both with temperature $T_i=1\,$keV (initial ion temperature), in a background of thermalized electrons with $T_e=500\,$keV (as extracted from the 3D simulation) predicts the mode with the maximum growth rate to be $k_{\rm max}\simeq 1.5\,\omega_L/c$ in good agreement with $k^{\mbox{\rm\tiny sim}}_{\mbox{{\rm\tiny IWI}}}$, and a corresponding growth rate $\Gamma(k^{\mbox{\rm\tiny sim}}_{\mbox{{\rm\tiny IWI}}})\simeq 0.11\,t_L^{-1}$~(see Appendix~\ref{app:IWI} which provides a generalization of Ref.~\cite{Grassi2017}). 
This theoretically predicted growth rate is larger than the one measured in the simulation.
This can be easily explained noting that this theoretical model considers symmetric and rather cold (1~keV) ion flows, 
whereas our 3D simulation evidences a quite large (with respect to the background ions) temperature 
$T^{\mbox{\rm\tiny sim}}_{\mbox{\tiny i,{ \rm HB}}}\simeq13.8$ keV for the HB reflected beam. 
This higher temperature can drastically reduce the maximum growth rate of the IWI.
Indeed, considering two counter-streaming beams with $T_{i}\simeq13.8$ keV, this fluid approach (confirmed by additional 2D3V PIC simulations, not shown) 
 predicts the instability to be completely quenched. 
In order to obtain the growth rate of the IWI in the presence of two different ion flow temperatures, we performed a complementary simulation in a reduced 2D3V geometry.
This simulation is initiated considering two overlapping and counter-streaming ion flows with temperatures $T_{i1}=1\,$keV and $T_{i2}=13.8\,$keV, respectively,
in a neutralizing electron background with zero drift velocity and temperature $T_e=500\,$keV (as extracted from the 3D simulation). 
This reduced simulation is found to lead to the development of an IWI with a growth rate $\simeq0.043\,t_L^{-1}$,
consistent with that obtained in the full 3D simulation, and thus confirming the dominant role of the IWI in the formation of the filamentary structures observed
in the 3D simulation, Fig.\ref{3DSpol45}b). 

At a later time, $t\simeq 65\:t_L$, the IWI-generated magnetic field in the 3D simulation reaches 
$B^{\mbox{\rm\tiny sim}}_z\simeq 3\:m_e\omega_Lc/e$ ($B^{\mbox{\rm\tiny sim}}_z\simeq 3\times10^8\,$G for $\lambda_L=1\,{\rm \mu m}$ ). 
Considering all the reflected ions in a region of extension $\simeq\lambda^{\mbox{\rm\tiny sim}}_{{\rm\tiny IWI}}$ to be confined in a cylindrical current filament with diameter $\simeq\lambda^{\mbox{\rm\tiny sim}}_{{\rm\tiny IWI}}/2$, the corresponding magnetic field would be $B=2\pi\lambda^{\mbox{\rm\tiny sim}}_{{\rm\tiny IWI}} en_0v_{\mbox{\tiny HB}}/c\simeq 8.5\,m_e\omega_Lc/e$, larger than that observed in the simulation.
Considering a partial screening of the ion currents by thermalized electrons (with $T_e=500\,$keV) following the model proposed in Ref.~\cite{AchterbergNL}, 
we expect the magnetic field at saturation to be $B_{\rm sat}\simeq4.1\,m_e\omega_Lc/e$, only slightly larger than the one measured in our simulation.
This indicates that, at the end of the 3D simulation, the IWI is close to saturation, with a significant part of the ions trapped in the filaments whose currents are partially screened by thermalized electrons.\\

%%% SHOCK FORMATION
\section{Ion Weibel-mediated collisionless shock formation}\label{sec:Shock}

The situation demonstrated in the 3D simulation presented in the previous Section is known to be the early stage of shock formation. 
To be able to reach shock formation in the simulation using a reasonable computation time, we performed a 2D3V simulation with an artificially reduced ion to electron mass ratio $m_i=100\,m_e$.
\begin{figure}
\includegraphics[width=0.49\textwidth]{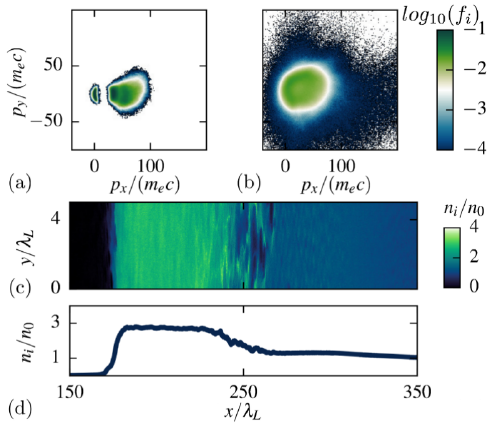}
\caption{\label{fig:IonPhaseSpaces}Radiation pressure driven collisionless shock formation in a 2D3V simulation with reduced ion mass ${m_i=100\,m_e}$.
a-b) Ion distribution in $p_x$-$p_y$ at $t=30\,t_L$ (before shock formation) and $t=515\,t_L$ (after shock formation), respectively.
c) Ion density $n_i(x,y)$ and d) averaged ion density $\langle n_i(x)\rangle_y$ at $t=515\,t_L$.}
\end{figure}
Reducing the ion mass while keeping all other parameters unchanged gives $v_{\mbox{\tiny HB}}=0.28\,c$, in good agreement with that measured in the 2D3V simulation $v^{\mbox{\rm\tiny sim}}_{\mbox{\tiny HB}}\simeq0.29\,c$.  
Figures~\ref{fig:IonPhaseSpaces}a-b show, at $30\,t_L$ and $515\,t_L$ respectively, the phase space $p_x$-$p_y$ of a region with extension $\simeq \lambda_L$ initially close to the surface, moving at $v^{\mbox{\rm\tiny sim}}_{\mbox{\tiny HB}}$. 
In Figure~\ref{fig:IonPhaseSpaces}a (before shock formation) we clearly identify the background plasma distribution centered around $p_x\simeq 0$ and the HB-reflected beam, centered around $p_x\simeq 71\,m_ec$, corresponding to $\sim 2v^{\mbox{\rm\tiny sim}}_{\mbox{\tiny HB}}$. 
Figure~\ref{fig:IonPhaseSpaces}b shows at $t=515\,t_L$ isotropization of the ion distribution function around the momentum $p_x\simeq 30 m_ec$ corresponding to $v_{\mbox{\tiny HB}}$. Full thermalization is not yet reached as the measured ion temperature $T^{\mbox{\rm\tiny sim}}_{i}\simeq 1.1\,m_ec^2$ is lower than $T_i=(\gamma_{\mbox{\tiny HB}}-1)m_ic^2 \simeq 4.5\,m_ec^2$ obtained considering that all the drift kinetic energy is dissipated into thermal energy. 
Nevertheless, a density jump (up to $3n_0$), consistent with the Rankine-Hugoniot (RH) prediction for a non-relativistic two-dimensional flow~\cite{RHconditions}, is observed in Figs.~\ref{fig:IonPhaseSpaces}c-d, suggesting that the shock is formed~\cite{RuyerPRL,RuyerPOP}. Furthermore, the shock front located around $x\simeq250\,\lambda_L$ in Figs.~\ref{fig:IonPhaseSpaces}c-d and with characteristic width $\simeq50\,\lambda_L$, corresponding to $\sim 50$ ion skin depths, moves with a velocity 
$v^{\mbox{\rm\tiny sim}}_{\mbox{\tiny sh}}=0.42\,c$, consistent with the RH prediction $v_{\mbox{\tiny sh}}\simeq 0.43\,c$.

Note that in this configuration the shock front is created deep inside the target, far from the laser-plasma interaction surface, and there is a clear distinction between the role of laser-plasma interaction processes on the one hand and the evolution of the IWI to a Weibel mediated collisionless shock on the other hand.\\

%%% SURFACE INSTABILITY
\section{Suppression of the surface instability}\label{sec:SurfInsta}

As we pointed out in Sec.~\ref{sec:IWI3dsim}, our study suggests that the optimal laser configuration to create fast quasi-neutral flows while keeping a low amount of laser-generated hot electrons is obtained using linear S-polarization at $\theta\sim 45^{\circ}$ incidence. 
One could expect the choice of circularly (C-) polarized laser pulse at normal incidence to be an even more favorable configuration as it is ideally expected to strongly reduce electron heating~\cite{Macchi05,Robinson09,Schlegel09}.
However, as will be shown in what follows, the use of C-polarized light at normal incidence leads to a strong electron heating due to the modulation of the interaction surface driven by surface electromagnetic instabilities.
The resulting fast electrons then propagate into the target, driving a return current, and the dynamics is mainly governed by electron instabilities, as observed in Refs.~\cite{FiuzaPRL,Ruyer2015} using P-polarization and normal incidence. 
Instead, we demonstrate, by means of 2D3V simulations, that the surface instability can be mitigated by a current generated along the surface when the target is irradiated at oblique incidence in linear polarization and the electron heating due to the laser-surface interaction is reduced.

The simulations are performed considering a similar configuration to the one discussed before: a laser with $a_0=70$ is irradiating a plasma with density $n_0=49\,n_c$.
The spatial resolution is here set to $\lambda_L/320$, $L_x=32\lambda_L$, $L_y\simeq4.3\lambda_L$ and 49 particles-per-cell are used for each species.
Simulations C-polarized laser at normal incidence, and S-polarized light at $\theta=45^{\circ}$ incidence are reported. As a reference, we also shown the case of P-polarized laser at normal incidence, corresponding to~\cite{FiuzaPRL,Ruyer2015}.

\begin{figure}
\includegraphics[width=0.48\textwidth]{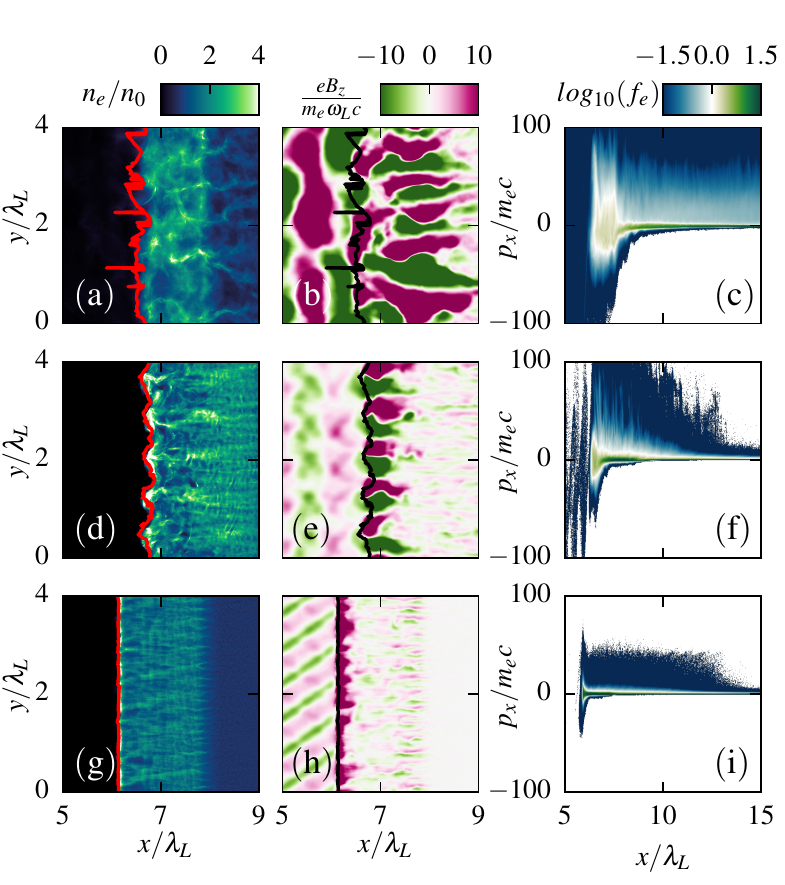}
\caption{\label{C00vsS45} Effect of the laser polarization and angle of incidence on the surface instability and electron heating. 2D3V simulations with P- and C-polarized laser pulses at normal incidence $\theta=0^{\circ}$ (top panels and middle panels, respectively) and S-polarized pulse at oblique incidence $\theta=45^{\circ}$ (bottom panels), at $t=25\,t_L$. a-d-g) Electron density. b-e-h) Magnetic field $B_z$ averaged on a laser period. c-f-i) $x$-$p_x$ electron phase space.
The red and black lines denote the position of the laser-plasma interaction surface.}
\end{figure}

In the case of C-polarized laser pulses, the electron density and the ion surface density profile, shown at $t=25\,t_L$ in Fig.~\ref{C00vsS45}d, evidence the generation of strong corrugations of the surface. 
Different types of instabilities have been proposed to explain these modulations~\cite{AdamHeron2006,Sgattoni15,Eliasson,Mori2016}, which lead to the production of a large amount of hot electrons that propagate at relativistic velocity in the target as shown in Fig.~\ref{C00vsS45}f. 
The correspondence of the surface rippling with the magnetic field structures, highlighted in Fig.~\ref{C00vsS45}e, suggests an electromagnetic nature of the instability at this stage. 
In order to avoid the formation of these structures, we varied the configuration of the interaction. Using S-polarization and $\theta=45^{\circ}$, the development of the surface magnetic structures is completely suppressed. 
The efficiency of this configuration in reducing the surface instability is clarified by Fig.~\ref{C00vsS45}g, where we show that the surface profile remains approximately flat and no magnetic filaments are present at the surface (Fig.~\ref{C00vsS45}h). 
Accordingly, there is an evident decrease of fast electron production with respect to the cases of P- or C-polarized light (Fig.~\ref{C00vsS45}c-f).

The suppression of the surface instability is linked to the establishment of a transverse slowly varying (with respect to the laser period) electron current $J_y$ at the surface.
The stabilizing role of this current is clear in a framework in which the surface magnetic structures are driven by electron Weibel-like instability 
for which one would expect filaments with magnetic field $B_z$ and wavevector $k_y$ to develop as in Fig.~\ref{C00vsS45}e.
A coherent motion of electrons along the $y$-direction, as evidenced by the surface current observed in the S-polarized $\theta=45^{\circ}$ simulation, prevents the confinement of the particles in a filament, removing the feedback mechanism for the instability growth.

The formation of similar current sheets at the interaction surface has been already observed in PIC simulations at non-normal incidence~\cite{Vshivkov1998,Mima2004}. This electron current produces a unipolar magnetic field at the surface, as observed in Fig.~\ref{C00vsS45}h, that reaches  at later times a strength comparable to that of the incident laser field.
A simplified model for the generation of the transverse current is now outlined.

For an electromagnetic plane wave with intensity $I(t)$ obliquely incident at an angle $\theta$ on the planar surface of a medium with reflectivity $R \leq 1$, the flow of electromagnetic momentum ${\bf P}$ transferred at the surface reads
\begin{equation}\label{eq:Pperppar}
(P_x,P_y)=\left((1+R)\frac{I}{c}\cos^2\theta \, , \,(1-R)\frac{I}{c}\sin\theta\cos\theta \right)
\end{equation}
where all quantities are expressed in the frame co-moving with the surface (primed notations have been dropped here).
The $P_x$ component corresponds to the standard radiation pressure on the surface (that which drives the Hole-Boring process) while $P_y$ describes the transfer of momentum to electrons in the direction parallel to the surface, and gives rise to a current in the skin layer. 
In turn, this current generates a magnetic field ($B_z$) and, by induction, an electric field ($E_y$) which counteracts the acceleration of electrons along the surface and, at late times, transfers part of the absorbed momentum to ions. 
To describe this process, we introduce the ponderomotive force in the skin layer ${\bf f}_p(x,t) \simeq {\bf P}\exp(-2x/\ell)/(\ell/2)$, where $\ell$ is an appropriate screening length for the laser electromagnetic fields, and we use cold fluid equations for the plasma electrons, yielding for the current
\begin{equation}
\ddt J_{y} = \frac{\omega_{pe}^2}{4\pi}E_{y}-\frac{e}{m_e}f_{py}
\end{equation}
coupled with Maxwell's equations (see Appendix~\ref{app:SurfInsta}). We also neglected the contribution of the ion current because of the large mass difference with the electrons. Solving for the electron current we obtain
\begin{equation}\label{Jye}
J_{y}= \frac{4}{3}\frac{e\omega_{pe}}{m_ec}\left(\mbox{e}^{-\omega_{pe}x/c}-2\mbox{e}^{-2\omega_{pe}x/c}\right)\times \int_0^t f_{py}(0,t')\mbox{d}t' \; ,
\end{equation}
where we have assumed $\ell \simeq c/\omega_{pe}$, neglecting relativistic corrections on the electromagnetic wave penetration. Within this assumption, and considering for simplicity a flat-top profile $I(t)=I_0\Theta(t)$, the maximum value of $J_{y}$ at time $t$ is
\begin{equation}
J_{y}^{\mbox{\rm\tiny max}} \simeq \frac{\pi}{12}a_0^2\frac{t}{t_L}\left(\frac{n_0}{n_c}\right)^{1/2}(1-R)\sin(2\theta_i)\,en_cc \; .
\end{equation}

Even for very small absorption ($1-R\lesssim10^{-2}$) the transverse current is large enough to stabilize the surface instability.  
Saturation of the growth of the current and the associated magnetic field $B_{z}$~(see Appendix~\ref{app:SurfInsta}) will eventually occur when the cyclotron frequency $\omega_c=\omega_{pe}$, i.e. when the electron gyroradius equates the skin depth.
\begin{figure}
\includegraphics[width=0.49\textwidth]{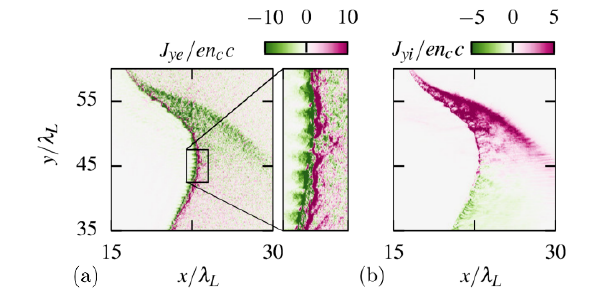}
\caption{\label{fig:WithProfile}{Mitigation of the surface instability by a surface current in a 2D3V simulation considering a finite laser spot size $10\,{\rm \mu m}$. Electron (a) and ion (b) current $J_{y}$ at $t= 85t_L$.}}
\end{figure}

In order to confirm that a finite laser spot size does not prevent the creation of the surface current and mitigation of the surface instability, we performed a 2D3V simulation considering a Gaussian transverse profile and focal spot ($1/e^2$ in intensity) $10\:{\rm \mu m}$. The simulation box has been enlarged to $64\,\lambda_L\times128\,\lambda_L$ with all other parameters unchanged. 
Also for a finite spot we observe the ambipolar electron current predicted by Eq.~\eqref{Jye}, that produces and confines the positive magnetic field in the skin layer, see Fig.~\ref{fig:WithProfile}a. The ion current, Fig.~\ref{fig:WithProfile}b, is directed along the positive $y$, i.e. in the direction predicted by momentum conservation, except in the region $y<45\,\lambda_L$, where the intensity gradient in the transverse direction due to the finite spot causes the local ponderomotive force to be in the negative $y$ direction.

%%% CONCLUSIONS
\section{Conclusions}\label{sec:conclusions}

In conclusion, we have demonstrated the possibility to efficiently drive the ion Weibel instability in the collisionless regime on UHI laser facilities. 
We have identified the optimal experimental configuration as linear S-polarization at oblique incidence. 
This configuration allows for the stabilization of the surface instability and in turn for the reduction of hot electron production. 
This situation, dominated by ion instabilities as in most astrophysical scenarios, is shown to potentially lead to Weibel-mediated collisionless shocks.   

%%% MISC.
\section*{Acknowledgements}
The authors thank L. Gremillet, A. Sgattoni, V.~Tikhonchuk and T. Vinci for valuable discussions and the {\sc Smilei} development team for technical support.
Financial support from Grant No. ANR-11-IDEX-0004-02 Plas@Par is acknowledged. 
AG acknowledges support from the Université Franco-Italienne through the Vinci program.
AG and AM acknowledge the PRIN project "Laser-Driven Shock Waves" (2012AY5LEL-002) sponsored by MIUR, Italy.
This work was performed using HPC resources from GENCI-TGCC (Grant 2017-x2016057678). 

%%% APPENDICES
\appendix

\section{Hole Boring velocity}\label{app:Piston}
We provide the extension to the usual calculation of the Hole-Boring (HB) velocity $v_{\mbox{\tiny HB}}$, for the case of a plane wave of intensity $I$ and frequency $\omega$, impinging at an angle $\theta$ on a perfectly reflecting target. 
We assume the plane wave wavevector to be $\vk=(k_x,k_y,0)$ and the plasma surface to move with velocity ${\bf v}_{\mbox{\tiny HB}}=v_{\mbox{\tiny HB}}\hat{\bf x}$, in the laboratory frame $L$.
Therefore, in the frame co-moving with the target surface $L'$, the incident wave wavevector becomes
\begin{eqnarray}
\nonumber k_x ' &=& \gamma(k_x-\beta \omega/c)= \gamma k(\cos\theta-\beta)\, , \\
\nonumber k_y' &=&k_y =k \sin\theta\,,
\end{eqnarray}
where $k = \omega/c$, $\beta = v_{\mbox{\tiny HB}}/c$, $\gamma = (1-\beta^2)^{-1/2}$.
In $L'$ the incidence angle $\theta'$ is thus given by
\begin{eqnarray}\label{ThetaPrime}
\tan\theta'= \frac{k_y'}{k_x'}=\frac{\sin\theta}{\gamma(\cos\theta-\beta)}\, .
\end{eqnarray}
In $L'$ a plane wave of intensity $I'$ interacts with an immobile plasma with incidence angle $\theta'$, the radiation pressure is thus given by:
\begin{eqnarray}
\nonumber P' = \frac{2I'}{c} \cos^2\theta'.
\end{eqnarray}
Exploiting the Lorentz's transformations for the wave electromagnetic fields, we obtain $I' = I(1-\beta)/(1+\beta)$. Since the pressure is a relativistic invariant $P'=P$, thus the radiation pressure $P$ in the frame $L$ reads
\begin{eqnarray}\label{RP}
\nonumber P \! =\!  \frac{2I}{c}\frac{1-\beta}{1+\beta}\frac{1}{1+\tan^2\theta'}=\frac{2I}{c}\frac{1-\beta}{1 + \beta}\!\left[1+\frac{\sin^2\theta}{\gamma^2(\cos(\theta)-\beta)^2}\right]^{\!-1}\!\!\!\!\!\!\!.
\end{eqnarray}
The HB velocity can now be estimated by balancing the laser radiation pressure with the flux of ion momentum $P_i=n_i \gamma\beta c(2m_i\gamma\beta c)$,~\cite{Schlegel09}.
Solving $v_{\mbox{\tiny HB}}=\beta c$ yields the Hole-Boring velocity as a function of $\theta$. An analytical solution can be obtained but is quite cumbersome, and the solution can be easily found numerically. \\
Note that from Eq.~\eqref{ThetaPrime} we obtain that for $\beta= \cos \theta$ the wave propagates parallel to the surface in the $L'$ frame, and $ c\cos \theta$ appears has a natural upper limit for the Hole-Boring velocity. \\   

\section{IWI growth rate}\label{app:IWI}

The IWI can be characterized within a relativistic warm fluid approach, as discussed in Ref.~\cite{Grassi2017}. 
We compute the instability growth rate $\Gamma$ starting from the relativistic fluid equations:
\begin{eqnarray}
\nonumber \partial_t n_{\alpha}+{\bf \nabla}\cdot\left( n_{\alpha}{\bf V}_{\alpha}\right) &=&0 \\
\nonumber h(\mu_{\alpha}) \left[ \partial_t {\bf P}_{\alpha}+\left( {\bf V}_{\alpha} \cdot {\bf \nabla}\right) {\bf P}_{\alpha}\right] &=& q_{\alpha}\left[ {\bf E} + \frac{ {\bf V}_{\alpha}}{c}\times{\bf B}\right] - \frac{ {\bf \nabla} \mathcal{P_{\alpha}} }{n_{\alpha}}
\end{eqnarray}
where $\alpha$ is the index of species, $\mu_{\alpha}=m_{\alpha}c^2/T_{\alpha}$ and $h =1+ \left(e+ \mathcal{P}\right)/(nmc^2)$ is the normalized enthalpy, with $e$ the internal energy and $\mathcal{P}$ the thermal pressure. \\
We consider two counter-streaming proton beams with velocity ${\bf V}=\pm V_0\hat{x}$, density $n_{\pm}=n_0/2$ and temperature $T_{\pm}=T_0$ [so that $\mu(T_{\pm})=\mu_0$]. The neutralizing background is provided by warm thermalized electron with density $n_0$ and temperature $T_e\gg T_0$, so that the fluid approach can not be used (being valid if the thermal velocity is much smaller than $\Gamma/k$). As demonstrated by means of a kinetic approach in~\cite{Lyubarsky2006}, the contribution of the electrons in the large temperature limit disappears from the dispersion relation. 
Linearizing the fluid equations for the ions coupled with Maxwell's equations, and looking for purely transverse unstable modes, we obtain 
\begin{widetext}
\begin{equation}\label{DispRel}
\frac{\omega^2}{c^2}-k_y^2-\frac{\omega_{pi}^2}{h(\mu_0)\gamma_0^3c^2} -\frac{\omega_{pi}^2}{h(\mu_0)\gamma_0c^2}\frac{k_y^2V_0^2}{\omega^2-T_0k_y^2/(m_ih(\mu_0)\gamma_0)}=0
\end{equation} 
\end{widetext}
where for simplicity we have assumed ${\bf k}=(0,k_y,0)$ and $\omega_{p\alpha}^2 = 4\pi n_0q_{\alpha}^2/m_{\alpha}$.
Equation~\eqref{DispRel} can be numerically solved for $\omega=i\Gamma$ with $\Gamma>0$ to obtain the IWI growth rate.

\section{Model for the electron transverse current}\label{app:SurfInsta}

The presence of the transverse current slowly varying (with respect to the laser period) in the skin layer is related to the absorption of electromagnetic (EM) momentum which can occur for oblique incidence at reflectivity $R<1$. 
Considering the test case of an EM plane wave with intensity $I=I(t)$ incident at an angle $\theta$ on the planar surface of the medium (filling the $x>0$ region) with $xy$ as the plane of incidence, the flow ${\bf P}$ of EM momentum through the surface has two components as given by Eq.~\eqref{eq:Pperppar}.
These components are obtained starting from Fresnel formulas for the laser EM fields at the surface, and calculating the flow of momentum at the surface using Maxwell's stress tensor.
The $P_{x}$ component corresponds to the radiation pressure on the surface (which drives the Hole-Boring process), while $P_{y}$ describes the transfer of momentum to electrons in the direction parallel to the surface, giving rise to a current in the skin layer. 
In turn, this current generates a magnetic field and, by induction, an electric field which counteracts the transverse acceleration of electrons and transfers part of the absorbed momentum to ions. Notice that in the case of a surface moving with the $v_{\mbox{\tiny HB}}$ velocity the formula above has to be considered in the co-moving frame. 

To describe this process, we introduce the ponderomotive force in the skin layer ${\bf f}_p \simeq {\bf P}\exp(-2x/\ell)/(\ell/2)$, where $\ell$ is an appropriate screening length for the laser EM field in the plasma, and use cold fluid equations for the electrons:
\begin{eqnarray}
\label{current}\ddt j_{sy} &=& \frac{\omega_{pe}^2}{4\pi}E_{sy}-\frac{e}{m_e}f_{py} \; , \\
\label{Maxw1}\ddx B_{sz} = -\frac{4\pi}{c}j_{sy} \; &,& \; \ddx E_{sy} = -\frac{1}{c}\ddt B_{sz} \; ,
\end{eqnarray}
where $\omega_{pe}^2=4\pi e^2n_0/m_e$ and the suffix ``$s$'' means that all fields are slowly varying on the temporal scale of the laser period, so that the displacement current is negligible. We have also neglected the contribution of the ion current to $B_{sz}$ because of the large mass difference with the electrons. 
Combining the previous equations, we obtain an inhomogeneous Helmoltz equation for the electric field component $E_{sy}$ 
\bea\label{Helmoltz}
\left(\ddx^2-\frac{\omega_{pe}^2}{c^2}\right)E_{sy}=-\frac{4\pi e}{m_ec^2}f_{py} \; .
\eea
The particular solution of Eq.~\eqref{Helmoltz} can be obtained as a Laplace transform in space:
\bea
\nonumber \hat{E}_{sy}(s,t)&=&-\frac{4\pi e/m_ec^2}{s^2-\omega_{pe}^2/c^2}\hat{f}_{py}(s,t), \\
\nonumber E_{sy}(x,t)&=&\frac{1}{2\pi i}\int_{a-i\infty}^{a+i\infty}\hat{E}_{sy}(s,t)\mbox{e}^{sx}\mbox{d}s \; ,
\eea
where $a$ is any length larger than the convergence abscissa, which is determined by the spatial profile of $f_{py}(x,t)$.
Considering $\ell \simeq c/\omega_{pe}$, consistent with relatively weak absorption in an highly overdense plasma, we obtain
\bea
\nonumber E_{sy}(x,t) \simeq \frac{f_{py}(0,t)}{3n_0e}\left(\frac{3}{2}(1-C)\mbox{e}^{-\omega_{pe}x/c}-\mbox{e}^{-2\omega_{pe} x/c}\right) \; .
\eea
where $C$ is a constant will be fixed from the boundary conditions.
From Eqs.~\eqref{current} and~\eqref{Maxw1}, we obtain for the magnetic field
\begin{equation}
B_{sz} \simeq \frac{4\omega_{pe}^2}{3en_0c}\left(\mbox{e}^{-\omega_{pe}x/c}-\mbox{e}^{-2\omega_{pe} x/c}\right)\int_0^tf_{py}(0,t')\mbox{d}t'\; ,
\end{equation}
where we have assumed at the laser-plasma surface $B_{zs}(x=0)=0$, so that $C=-1/3$. Proceeding similarly, we obtain for the electron current Eq.~\eqref{Jye}.\\
The maximum value of the magnetic field is found at $x=(c/\omega_{pe})\ln 2$, and considering a flat-top profile $I(t)=I_0\Theta(t)$ [with $\Theta(t)$ the Heaviside function], it reads 
\bea
B^{\mbox{\tiny max}}_{sz} \simeq \frac{\pi}{6}a_0^2\frac{t}{t_L}(1-R)\sin(2\theta_i)\frac{m_e\omega_Lc}{e} \; .
\eea 
Notice that $E_{sy}(x,t)>0$ and $B_{sz}(x,t)>0$, consistent with the simulation results.

\bibliography{Biblio}

\end{document}